# Providing Intermodal Route Alternatives

Matthias Prandtstetter a, Clovis Seragiotto a, Markus Straub a, Babis Magoutas b, Efthimios Bothos b, Luka Bradesko c*

*[a]AIT Austrian Institute of Technology GmbH*
*Giefinggasse 2, 1210 Vienna, Austria*
*[b]Institute of Communication and Computer Systems (ICCS), National Technical University of Athens (NTUA)*
*9 Iroon Polytechniou street, Zografou, 15780, Athens, Greece*
*[c]Institut Jozef Stefan*
*Jamova cesta 39, 1000 Ljubljana, Slovenia*

**Abstract**

Within this paper, we present a novel routing algorithm capable of providing not only truly intermodal routes but also coming up with route alternatives. These route alternatives feature different route and mode choices while still optimizing the same objective function (e.g. travel time). We therefore, provide a first presentation of the next generation routing service, which are fundamental for the introduction of Mobility-as-a-Service in the passenger sector or synchromodality in the freight transportation sector. We finally provide a showcase of motorhome routing where the full potential of the presented routing algorithm is shown.

---

\* Corresponding author. Tel.: +43 50550 6692; fax: +43 50550 6439.
  *E-mail address:* matthias.prandtstetter@ait.ac.at

## 1. Introduction

During the last few years, our mobility behavior was strongly influenced by the technological developments. On the one hand, the internet and web services made it possible that we started planning our trips not by using (unhandy) paper maps but by relying on computer algorithms providing us with route suggestions. On the other hand, mobile phone technology provided us with the possibility to do this planning not only at home, i.e. pre-trip, but we are now able to re-plan our trip even during traveling. This resulted in the fact that we, nowadays, strongly depend on powerful routing services. In this paper, we therefore present the so-to-say next generation routing service which is capable of not only providing multimodal or "bike&ride" trips but also of coming up with truly intermodal routes, suggesting in addition to where to travel, also how to travel. These applications build a fundamental service when thinking about next-to-come applications like synchromodality or Mobility-as-a-Service.

*1.1. Structure*

The remainder of the paper is structured as follows: First, we give a detailed description of state-of-the-art route planning services including a suggestion on how to extend them towards the next generation routing services. Then, we present a novel approach on how to generate sets of route alternatives according to user preferences. Finally, we present a motorhome routing showcase. A summary including future work concludes the paper.

## 2. Intermodal route planning

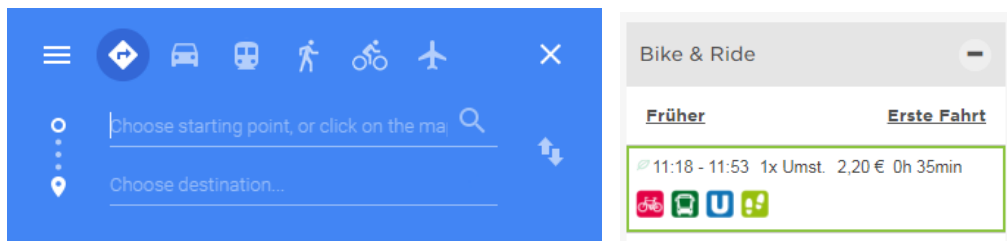

Figure 1 the well-known route planning interface as provided by Google Maps (2017; left) and an intermodal route planning interface as provided by AnachB.at (2017; right)

For simplicity, we focus on passenger transportation only within this paper. However, the concepts described apply to freight transportation as well, although some of the terms used differ in the freight transportation context.
When speaking of *intermodality* (in the context of passenger transportation), one normally refers to the fact that *along one individual trip multiple* (but at least two) *different modes* of transportation are used. In order to avoid confusion, we want to mention that *multimodality* only refers to the possibility *to choose* among multiple modes of transportation. E.g., for a trip to visit a theater, there are multiple options to get there: taking the taxi, driving the car or utilizing public transportation. One specific option might contain of taking the car towards a parking lot close to the theater and, after parking car, the theatergoer walks a few hundred meters. The modes utilized are, obviously, driving a car and walking. Therefore, we say that this trip is intermodal.
In many situations, users utilize route planning services for finding their fastest (or shortest) path from an origin location (also referred to as A in the remainder of this paper) towards a destination location (also referred to as B in the remainder of the paper). Most of the route planning services available in the internet are multimodal (cf. Figure 1 (left)), while only a few of them provide intermodal functionalities (cf. Figure 1 (right)). E.g., when planning a route with Google Maps (2017), the user provide her origin and destination as well as the desired mode of transportation. The request results in a proposition of one or more routes from A to B utilizing the selected mode of transportation. With the intermodal service of AnachB.at (2017), the user provides her origin and destination as well as the two desired modes of transportation. For the service shown in Figure 1 (right), the two selected modes are riding a bike and public transport. Therefore, the request results in one or more routes utilizing for the first leg a bike and for the second (last) leg the public transport. Please, be aware that one does not consider public transport intermodal in a broader context, although normally consisting of more than one legs and in many cases incorporating even more than one modes of transportation like bus, tram, train and/or walking.

## 2.1. Shortcomings of existing multimodal/intermodal route planning services

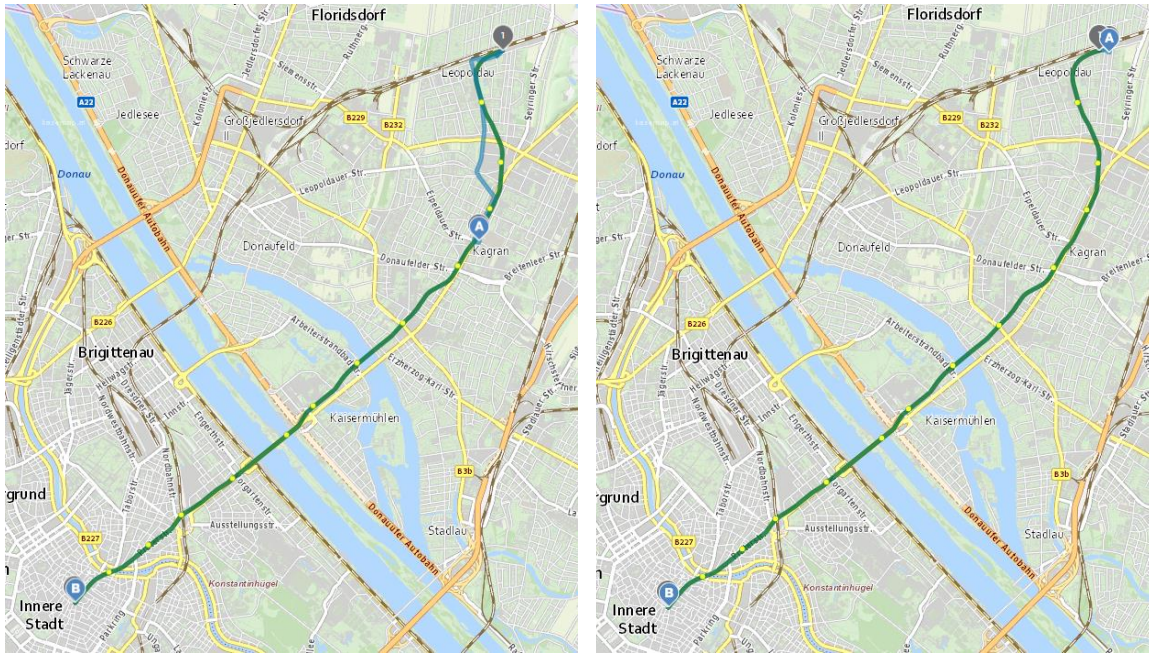

Figure 2 bike&ride routes as provided by the intermodal route planning service AnachB.at (2017). While on the left side a public transport station far off the expected travel direction is proposed (location 1 at the top of the picture), the distance traveled by bike is 60m on the route proposed on the right side.

Although the internet provides access to multimodal (e.g. Google Maps (2017)) and (a few) intermodal route planning services (e.g. AnachB.at 2017), all of them encompass significant shortcomings. For multimodal route planning services, the main drawback is that they are not capable of providing intermodal routes, i.e. routes comprising a change among different modes of transportation along a trip. Obviously, this is, especially in the suburban-urban context, a major shortcoming as in the greater area of (big) cities it is often convenient to start a trip towards the city center with driving the car and then switching to public transport as soon as a high performance line (e.g. a suburban train or even an underground train) is met. This strategy is reasonable especially when living in a suburban area with very limited access to public transport.

Intermodal route planning services, on the other hand, suffer from the fact that they are normally designed in such a way that a change among different modes of transportation is mandatory. For the use case described in the previous paragraph, this is obviously not limiting but the desired behavior. However, when e.g. selecting "bike&ride", i.e. starting your journey with the bike and switching at a convenient station to the public transport service, it may happen that ridiculous routes are generated since intermodality *is forced*. In Figure 2, we present the results obtained via one of the official intermodal route planning services in Vienna (AnachB.at 2017). While for the first route, the routing service proposes to take the bike for a long distance (approx. 3.5km) in the "wrong" direction in order to change to public transport (most probably because there is no closer, appropriate bicycle parking identified), the routing service suggests to use the bike for only 60m in the second example. In both examples, we have chosen the origin very close to a public transport service. I.e., we would expect that in both cases, a routing service suggest that only the public transport is used.

## 2.2. The holistic approach

In order to overcome the above-mentioned shortcomings, we suggest to provide a *truly intermodal* route planning. This means that an ideal routing planning service comes up with an optimized route. Optimality in this context is, however, to also include sanity checks into the route planning, which guarantee that proposed routes are plausible. Furthermore, we suggest that mode choice decisions are not fixed by the user during the routing request but suggested by the routing service as part of the routing response. That is, as sketched in Figure 3, the user preselects the (for her) available modes of transportation. The routing service chooses then, among all possibilities, which is the most appropriate one. Obviously, the routing service evaluates all route options according to the chosen objective function (e.g. travel time, distance or any other convenient function like costs). Furthermore, in many situations, the user does not have her bike and/or car with her but e.g. due to a bike&ride trip earlier that day has the bike and/or car parked at a specific location. It is then convenient for the user if she can provide the current location of her vehicles, which the routing service considers during route planning. That way, the suggested routes can significantly differ from what you would expect from a state-of-the-art route planning service currently available in the World Wide Web.

Figure 3 A suggestion for a possible user interface. The user can provide origin and destination (including a desired departure or arrival time), the current location of her bike and/or car, and a desired objective function (e.g. travel time or distance). In addition, the user preselects those modes of transportations which would be acceptable (e.g. walking and public transport).

Finally, we want to claim that we expect that even the best routing service does not fit all user requirements. This is, among others, due to the fact that some route suggestions which are appropriate for one day are absolutely inappropriate on other days, e.g. due to differing surrounding parameters like appointments or weather conditions. We therefore suggest that a routing request does not end up with only one suggested route but with a set of alternative routes. However, we want to highlight that (especially in the context of public transport) varying the departure time (by a few minutes) and suggesting the then resulting routes as alternatives is not what we are aiming at. Instead, we suggest that alternatives are (also) computed via varying the utilized modes of transportation. For this purpose, we introduce the notion of weighting factors which are explained in more details in the next section.

**3. Generating route alternatives and persuading users towards more sustainable mobility**

When computing a shortest route in a given (street) network, it is convenient to employ the well-known Dijkstra's algorithm (Dijkstra 1959). The basic idea of this algorithm is twofold: On the one hand, the algorithm follows the basic dynamic programming principle (Bellman 1959) of dividing the original problem into subproblems, which can be solved to optimality and whose optimal solution can be utilized for generating an optimal solution for the original problem. Second, the algorithm operates on a (network) graph assuming that only non-negative edge weights are existing. This is, in most cases, naturally given (e.g. travel times or distances cannot be negative). However, in some situations (e.g. energy consumption) the weights of edges can be negative. In such cases, specific strategies have to be applied such that Dijkstra's algorithm is still coming up with meaningful results. We refer to Prandtstetter et al. (2013) for some details on this.

When aiming at intermodal routes, between two main approaches can be chosen: Either, one preselects possible points of interchange (e.g. appropriate bike&ride stations; referred to as X in the further context) and then computes routes from A to X and from X to B. Among all possible Xs, that one is chosen for which the sum of two routes (A to X, X to B) is minimal. Without having deeper knowledge, we assume that AnachB.at (2017) applies this approach based on the observed behavior as shown in Figure 2. The second, more complex yet preferable approach is to integrate all available modes of transportation into one large layered routing graph, where each layer corresponds to one mode of transportation. Interchanges between modes of transportations are only possible at appropriate locations (e.g. bike&ride stations), which are modelled as an extra edge connecting the corresponding layers in the routing graph. We refer to Prandtstetter et al. (2013) for more details.

In this paper, we followed the second, more complex approach but instead of explicitly modelling the multi-layered routing graph, we decided to implicitly model it by building one routing graph consisting of all (existing) roads and street links. We annotate each edge in this graph with the modes of transportation allowed to traverse that edge. Although the final implementation of Dijkstra's algorithm has to be adapted such that this implicit multimodality is respected, the memory usage of the algorithm can be significantly reduced. Further acceleration techniques such as Contraction Hierarchies (Geisberger 2008) can be applied. Please note, however, that dynamic edge weights (e.g. travel times based on current traffic state) cannot be realized when utilizing this acceleration approach.

Although all of the above mentioned techniques are necessary in order to come up with a reliable intermodal routing service, it generates only one (namely the best) route. There are, however, situations where the best route according to an arbitrarily chosen objective function is not the optimal one with respect to the understanding of the current user. For this purpose, it is convenient to come up with not only one suggestion but to provide a set of (route) alternatives to the user. One way to achieve this would be to alter the objective function. E.g., instead of computing only the fastest route, one could also compute the shortest one, the most energy efficient one and the one with the least number of left turns (just to mention some other possible objective functions). However, two main drawbacks can be identified with this approach: On the one hand, it might very easily happen that all of these routes incorporate the same mode of transportation (e.g. walking) which is very likely since in most cases there are many shortcuts (especially in cities) where access for modes of transportation other than walking is forbidden. Furthermore, energy consumption is, by nature, less for walking than for driving a bike or car. On the other hand, routes without left turns (just to pick one of the previously mentioned objective functions) might be rather unhandy and/or extremely slow compared to other routes.

We therefore suggest not to change the objective function but to systematically change the individual edge weights. For this purpose, we fix the objective function to computing the fastest route. However, the perception of travel time is dependent on many factors like travel speed, mode of transportation, level of service (e.g. stop&go vs. free-floating traffic), etc. Therefore, we suggest to introduce for each mode of transportation a multiplier between 1 and 100 which corresponds to the individual travel time perception. The routing algorithm is now adapted such that instead of summing up the actual travel times, the new edge weights to be utilized are the original weights multiplied with the individual multiplier.

As example: A factor of 2 for walking and 3 for biking would indicate that the corresponding user perceives 2 minutes on the bike just as 3 minutes walking. This means that the user would prefer a walk of 15 minutes over a bicycle ride of 11 minutes as 15x2=30 (walking) and 11*3=33 (bicycle). We want to highlight that in the case that all multipliers are set to identical values (e.g. 5) the same route result will be generated as without multipliers.

Based on the above presented approach, it is now possible to generate a set of route alternatives, which are, however, still trying to minimize the travel time, by simply coming up with different sets of multipliers. One could go even one step further by adjusting the multipliers according to the feedback of users, i.e. to personalize the routing service.

Nevertheless, we as a society are not only aiming to have choices among which we choose but we hope that on a holistic view more sustainable options are chosen more often than not so sustainable ones. We therefore suggest that in addition to the generation of routing alternatives, persuasive technologies, i.e. technologies designed to change attitudes or behaviours of the users through persuasion and social influence, but not through coercion (Fogg, 2002), are applied in order to positively influence users of route planning devices towards selecting environmental friendly routes. This does not mean that we are aiming at hiding options from users but at showing them options and showing them that e.g. other users were much more environmentally friendly than they are currently. For a detailed discussion on that topic, we refer to Anagnostopoulou et al. (2017a, 2017b).

Finally, we want to highlight that none of these functionalities is providing an added value if real-time data is not considered. For example, if a car-sharing system is operated in a city, then the current positions and availabilities of the vehicles is of rudimentary importance as otherwise a route planning involving the service is of no further use. Furthermore, current travel times (e.g. due to traffic congestion) have a major impact on route planning and need to be considered as otherwise the applicability of the route service is limited to a rather narrow user group not interested in real-time information.

*3.1. The motorhome routing*

Within this section, we want to introduce a further novel functionality which has not been considered so far in real-world applications. Motorhome users have a quite interesting routing behavior as they (obviously) travel the main distance with their motorhome but often want to change to other modes of transportations as soon as they are close to their target. This is either because the motorhome is parked at a campground (where it is connected to amenities) or because the motorhome is rather unhandy when e.g. entering historic cities with narrow streets. Therefore, a special case of intermodal route planning has to be provided where the first leg of a trip is always with a motorhome but the rest of the trip is optimized according to other parameters (like available large parking places).

For the use case of traveling to cities, we therefore suggest to provide a routing service where not only one route is provided but a set of route alternatives such that

- one route consists of motorhome usage for the long haul, a suggestion for parking at designated motorhome parking lots and further travelling by other modes of transportation.
- one route consists of car usage for the long haul, a suggestion for parking at a designated car parking lot and further travelling by other modes of transportation.
- one route consists of car usage with car parking as close to the destination as possible.

First, it is important to have specific motorhome routes, as motorhomes are in general broader or longer than conventional cars meaning that they cannot access all streets accessible by private cars. Further, parking is in some situations much harder for them.

Nevertheless, if a motorhome user wants to risk and follow the route for a conventional car, than the user should be presented with that option as well. In many cases, designated car parking lots are also large enough to handle motorhomes although not particularly indicated. The final option is for the "brave ones", i.e. drivers who think (or know) that they can enter the city with the motorhome even if parking lots and streets are not specifically designated for motorhomes.

Using this three-option approach, one can even overcome a further very important topic, which, unfortunately, is not yet solved to full satisfaction – namely: map data. Although there are services like OpenStreetMap, which provide map data for all over the world, the data quality is rather cumbersome in some regions. Especially when relying on such open source data special cases like motorhomes are only suboptimally covered.

We refer to the OPTIMUM project (2017) for further details including a demo app.

## 4. Conclusions and Future Work

Within this paper, we elaborated on the computation of *truly intermodal* routes (compared to multimodal routes and intermodal routes which enforce the usage of two specific modes of transportation). The main idea followed is that, instead of fixing the mode choice before route computation (as done in various state-of-the-art routing services available in the internet), we ask the user only to provide a set of *possible* modes of transportation, i.e. a set of modes of transportation, which can be utilized by the user e.g. due to the existence of a driving license or the ownership of a monthly ticket for the public transport. The routing service then comes up with suggestions on how to travel from A to B and which modes of transportation to use.

In order to further enhance the routing service we come up with a novel method on how to compute alternative yet realistic routes considering also preferences by the users. Finally, we presented a novel routing use case – motorhome routing – which is of high relevance in the field of tourism. All options previously presented are integrated in this use case and show their full potential.

For future work, we want to highlight that the adaption of the multipliers as proposed in the previous sections needs to be improved such that user specific routes can be generated. When adapting these multipliers one must not, however, forget that overfitting is not expedient as in that case hardcore car drivers will always get car routes leading to the fact, that they are never presented with alternatives. Therefore, it will be necessary to always provide also some route alternatives which do not perfectly fit the user preferences in order to have some room for persuasion towards more enviromental friendly routes.

## Acknowledgements

Research reported in this paper has been partially funded by the European Commission project OPTIMUM (H2020 grant agreement no. 636160-2). We thankfully acknowledge all partners within the project for their valuable contributions and discussions during project meetings.